\documentclass[sigconf]{acmart}
\AtBeginDocument{%
  }

\copyrightyear{2025}
\acmYear{2025}
\setcopyright{acmlicensed}\acmConference[AIQAM '25]{Proceedings of the 2nd ACM Workshop in AI-powered Question \& Answering Systems}{October 27--28, 2025}{Dublin, Ireland}
\acmBooktitle{Proceedings of the 2nd ACM Workshop in AI-powered Question \& Answering Systems (AIQAM '25), October 27--28, 2025, Dublin, Ireland}
\acmDOI{10.1145/3746274.3760393}
\acmISBN{979-8-4007-2056-7/2025/10}

\usepackage[capitalise,noabbrev,nameinlink]{cleveref}
\usepackage{soul}
\setstcolor{red}
\usepackage{tikz}
\usepackage{longtable}
\usepackage{tablefootnote}
\usepackage[edges]{forest}
\definecolor{hidden-draw}{RGB}{20,68,106}
\definecolor{hidden-pink}{RGB}{255,245,247}
\definecolor{red}{RGB}{255,0,0}

\usepackage[draft,textsize=footnotesize,textwidth=15mm]{todonotes}

\definecolor{paired-light-blue}{RGB}{198, 219, 239}
\definecolor{paired-dark-blue}{RGB}{49, 130, 188}
\definecolor{paired-light-orange}{RGB}{251, 208, 162}
\definecolor{paired-dark-orange}{RGB}{230, 85, 12}
\definecolor{paired-light-green}{RGB}{199, 233, 193}
\definecolor{paired-dark-green}{RGB}{49, 163, 83}
\definecolor{paired-light-purple}{RGB}{218, 218, 235}
\definecolor{paired-dark-purple}{RGB}{117, 107, 176}
\definecolor{paired-light-gray}{RGB}{217, 217, 217}
\definecolor{paired-dark-gray}{RGB}{99, 99, 99}
\definecolor{paired-light-pink}{RGB}{222, 158, 214}
\definecolor{paired-dark-pink}{RGB}{123, 65, 115}
\definecolor{paired-light-red}{RGB}{231, 150, 156}
\definecolor{paired-dark-red}{RGB}{131, 60, 56}
\definecolor{paired-light-yellow}{RGB}{231, 204, 149}
\definecolor{paired-dark-yellow}{RGB}{141, 109, 49}

\definecolor{bg1}{HTML}{FF9966}
\definecolor{bg2}{HTML}{CCE5FF}
\definecolor{bg3}{HTML}{FFCC99}
\definecolor{bg4}{HTML}{FFC107}
\definecolor{bg5}{HTML}{FFCCCC}
\definecolor{bg6}{HTML}{D5E8D4}
\definecolor{bg7}{HTML}{eeeeee}
\definecolor{bg8}{HTML}{cdeb8b}
\definecolor{bg9}{HTML}{dae8fc}
\definecolor{bg10}{HTML}{a2e6eb}

\definecolor{bg31}{HTML}{FFCDD2} 

\definecolor{bg32}{HTML}{F8BBD0}

\definecolor{bg33}{HTML}{E1BEE7} 

\definecolor{bg34}{HTML}{D7CCC8} 

\definecolor{bg35}{HTML}{B2DFDB} 

\definecolor{bg36}{HTML}{A5D6A7} 

\definecolor{bg37}{HTML}{FFF9C4} 

\definecolor{bg38}{HTML}{FFECB3} 

\definecolor{bg111}{HTML}{CB6843}

\definecolor{bg112}{HTML}{D77C5C}

\definecolor{bg113}{HTML}{E28E6E}
\definecolor{bg114}{HTML}{E89F7D}
\definecolor{bg115}{HTML}{EDAE8A}
\definecolor{bg116}{HTML}{F0BA95}
\definecolor{bg117}{HTML}{F3C29F}
\definecolor{bg118}{HTML}{F6CCAA}
\definecolor{bg119}{HTML}{F8D5B3}
\definecolor{bg120}{HTML}{FADCBD}
\definecolor{bg121}{HTML}{FCE6C7}

\definecolor{bg39}{HTML}{FFE0B2} 

\definecolor{bg40}{HTML}{3CB371} 

\definecolor{bg43}{HTML}{ffe5d9}

\definecolor{bg15}{HTML}{7FFFD4}

\definecolor{bg17}{HTML}{F0FFFF}

\definecolor{bg18}{HTML}{F5FFFA}

\definecolor{bg19}{HTML}{F8F8FF}

\definecolor{bg20}{HTML}{FFFFFF}

\definecolor{bg21}{HTML}{E1F5FE}

\definecolor{bg22}{HTML}{B3E5FC}

\definecolor{bg23}{HTML}{81D4FA}

\definecolor{bg24}{HTML}{4FC3F7}

\definecolor{bg25}{HTML}{29B6F6}

\definecolor{bg26}{HTML}{03A9F4}

\definecolor{bg27}{HTML}{039BE5}

\definecolor{bg28}{HTML}{0288D1}

\definecolor{bg29}{HTML}{0277BD}

\definecolor{bg30}{HTML}{01579B}

\definecolor{bg16}{HTML}{FFCC99}

\definecolor{pg51}{HTML}{E8F5E9} 
\definecolor{pg52}{HTML}{C8E6C9} 
\definecolor{pg53}{HTML}{B9F6CA} 
\definecolor{pg54}{HTML}{A9DFBF} 
\definecolor{pg55}{HTML}{BCF5A6} 

\definecolor{pg56}{HTML}{BEF1CE} 
\definecolor{pg57}{HTML}{CEF6EC} 
\definecolor{pg58}{HTML}{B7F0B1} 
\definecolor{pg59}{HTML}{B1F2B5} 
\definecolor{pg60}{HTML}{9DF3C4} 

\definecolor{pg61}{HTML}{DEF7E0} 
\definecolor{pg62}{HTML}{E8F8DC} 

\definecolor{pg63}{HTML}{EBF7E7} 
\definecolor{pg64}{HTML}{F0FDF4} 

\definecolor{pg65}{HTML}{F1FEE7} 
\definecolor{pg66}{HTML}{F7FFF6} 
\definecolor{pg67}{HTML}{FCFFE7} 
\definecolor{pg68}{HTML}{F4FFD2} 
\definecolor{pg69}{HTML}{EEFFE2} 
\definecolor{pg70}{HTML}{E3FDF5} 

\definecolor{connect-color}{RGB}{0,0,0}
\definecolor{middle-color}{RGB}{255,255,255}
\definecolor{leaf-color}{RGB}{173,216,230}
\definecolor{line-color}{RGB}{25,25,112}

\begin{document}

\title{Multimedia-Aware Question Answering: A Review of Retrieval and Cross-Modal Reasoning Architectures}

\author{Rahul Raja}
\authornote{Work does not relate to position at LinkedIn}
\email{rahul.110392@gmail.com}
\orcid{0009-0008-6744-7145}
\affiliation{%
  \institution{LinkedIn}
  \city{Sunnyvale}
  \state{CA}
  \country{USA}}
\affiliation{
\institution{Carnegie Mellon University}
  \city{Pittsburgh}
  \state{PA}
  \country{USA}}
  
\author{Arpita Vats}
\authornotemark[1]
\email{arpita.vats09@gmail.com}
\orcid{0009-0009-4831-4109}
\affiliation{%
 \institution{LinkedIn}
  \city{Sunnyvale}
  \state{CA}
  \country{USA}}
\affiliation{%
  \institution{Boston University}
  \city{Boston}
  \state{MA}
  \country{USA}}








\begin{abstract}
 Question Answering (Q\&A) systems have traditionally relied on structured text data, but the rapid growth of multimedia content—images, audio, video, and structured metadata has introduced new challenges and opportunities for retrieval-augmented QA. In this survey, we review recent advancements in Q\&A systems that integrate multimedia retrieval pipelines, focusing on architectures that align vision, language, and audio modalities with user queries. We categorize approaches based on retrieval methods, fusion techniques, and answer generation strategies, and analyze benchmark datasets, evaluation protocols, and performance tradeoffs. Furthermore, we highlight key challenges such as cross modal alignment, latency accuracy tradeoffs, and semantic grounding, and outline open problems and future research directions for building more robust and context-aware Q\&A systems leveraging multimedia data.
\end{abstract}



\maketitle

\section{Introduction}
Traditional Question Answering (QA) systems have primarily relied on textual data to extract or generate answers~\cite{farea2022evaluationquestionansweringsystems}. However, as user queries increasingly demand richer context and deeper understanding, there has been a significant shift toward incorporating multimedia data such as images, videos, audio, and structured metadata—into the QA pipeline ~\cite{9350580,gao2024retrievalaugmentedgenerationlargelanguage}. This evolution is fueled by the rise of large scale multimodal datasets and powerful pretrained vision language models that enable semantic understanding across modalities ~\cite{liang2024comprehensivesurveyguidemultimodal}.

Multimedia retrieval based QA systems aim to bridge the gap between textual queries and non-textual content by retrieving relevant multimodal evidence and reasoning over it to generate accurate, grounded responses ~\cite{hoang2025pdfretrievalaugmentedquestion}. These systems play a crucial role in diverse applications, including visual question answering (VQA), video QA, instructional QA, and retrieval-augmented generation (RAG) for multimedia content ~\cite{zheng2025retrievalaugmentedgenerationunderstanding}.

The combination of retrieval techniques such as sparse or dense indexing and approximate nearest neighbor (ANN) ~\cite{andoni2018approximatenearestneighborsearch} search with generative ~\cite{pang2025generativeretrievalalignmentmodel} models (e.g., transformers, LLMs) has led to robust QA architectures capable of handling complex queries that require spatial, temporal, or semantic inference across different data types~\cite{Saquete_2009}.
\begin{figure*}[ht!]
  \centering
  \resizebox{\textwidth}{!}{
    \begin{forest}
      forked edges,
      for tree={
        grow=east,
        reversed=true,
        anchor=base west,
        parent anchor=east,
        child anchor=west,
        base=center,
        font=\small,
        rectangle,
        draw,
        rounded corners,
        align=center,
        text centered,
        minimum width=5em,
        edge+={darkgray, line width=1pt},
        s sep=3pt,
        inner xsep=2pt,
        inner ysep=3pt,
        line width=0.8pt,
      },
      [
        \textbf{Multimedia QA Taxonomy and Strategies}, fill=orange
        [
          \textbf{Modality-Specific QA Systems}, fill=cyan, text width=20em
          [\textbf{Unimodal Language QA  }\cite{karpukhin2020densepassageretrievalopendomain, izacard2021leveragingpassageretrievalgenerative, borgeaud2022improvinglanguagemodelsretrieving, santhanam2022colbertv2effectiveefficientretrieval}, fill=cyan!30, text width=20em]
          [\textbf{Static Vision-Language QA}\cite{tan2019lxmertlearningcrossmodalityencoder, chen2020uniteruniversalimagetextrepresentation, zhang2021vinvlrevisitingvisualrepresentations, alayrac2022flamingo}, fill=cyan!30, text width=20em]
          [\textbf{Spatiotemporal Vision-Language QA}\cite{lei2019tvqalocalizedcompositionalvideo, sun2019videobertjointmodelvideo, yang2022zeroshotvideoquestionanswering}, fill=cyan!30, text width=20em]
          [\textbf{Acoustic-Language QA }\cite{li2018spokensquadstudymitigating,li2022learninganswerquestionsdynamic,abdelnour2018cleardatasetcompositionallanguage,baevski2020wav2vec20frameworkselfsupervised,radford2022robustspeechrecognitionlargescale}, fill=cyan!30, text width=20em]
        ]
        [
          \textbf{Task-Oriented Taxonomy}, fill=paired-dark-green, text width=20em
          [\textbf{Modality-Aware Entity QA }\cite{zhu2016visual7w, antol2015vqa}, fill=green!35, text width=20em]
          [\textbf{Causal Reasoning QA }\cite{zellers2019vcr, zellers2019hellaswag}, fill=green!35, text width=20em]
          [\textbf{Contextual Interaction QA }\cite{reddy2019coqa,choi2018quac,yang2022zeroshotvideoquestionanswering}, fill=green!35, text width=20em]
          [\textbf{Temporal Event QA }\cite{lei2020tvqa+}, fill=green!35, text width=20em]
          [\textbf{Cross-Modal Reasoning QA }\cite{alayrac2022flamingo, yang2021justask}, fill=green!35, text width=20em]
        ]
        [
          \textbf{Multimodal Retrieval Strategies }, fill=paired-dark-blue, text width=20em
          [\textbf{Dense Retrieval}\cite{akbari2021vatttransformersmultimodalselfsupervised, ni2021largedualencodersgeneralizable,qu2021rocketqaoptimizedtrainingapproach}, fill=blue!35, text width=20em]
          [\textbf{Multimodal Embedding Retrieval }\cite{radford2021learning, li2022blip, alayrac2022flamingo}, fill=blue!35, text width=20em]
          [\textbf{Cross-Modal Retrieval }\cite{akbari2021vatttransformersmultimodalselfsupervised, gabeur2020multi}, fill=blue!35, text width=20em]
          [\textbf{Temporal Video Segment Retrieval }\cite{li2020hero, lei2021less}, fill=blue!35, text width=20em]
          [\textbf{Audio-Visual Retrieval }\cite{korbar2018cooperative, morgado2020audio}, fill=blue!35, text width=20em]
        ]
      ]
    \end{forest}
  }
\caption{A hierarchical taxonomy of Multimedia QA systems categorized by input modality, task formulation, and retrieval strategy, highlighting key models and reasoning approaches across heterogeneous data types.}

  \label{fig:llm_challenges_forest}
\end{figure*}
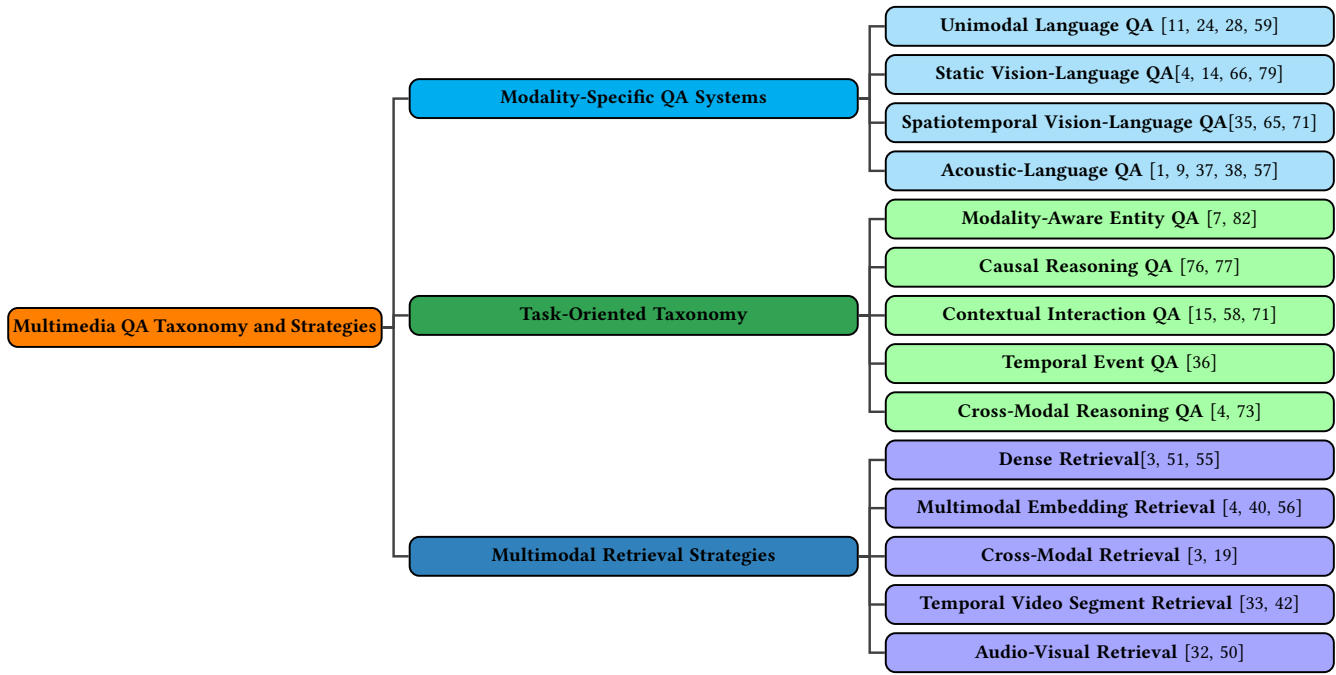
In this paper, we present a structured and focused review of QA systems that integrate multimedia retrieval capabilities. We categorize recent developments based on five key dimensions: modality specific QA systems, multimodal retrieval augmented architectures, temporal and spatial alignment strategies, knowledge-enhanced retrieval, and evaluation frameworks.Our goal is to offer a compact yet comprehensive guide for researchers and practitioners building next generation QA systems that operate over complex multimedia content. To facilitate a structured understanding of recent advancements, we present a hierarchical taxonomy of Multimedia QA systems, categorizing them by modality, task formulation, and retrieval strategy (Figure~\ref{fig:llm_challenges_forest}).

\section{Taxonomy of Multimedia QA Systems}
\label{sec:taxanomy}
Multimedia QA systems vary in how they process, fuse, and reason over inputs like text, images, audio, and video. This section presents a taxonomy based on input modalities, reasoning depth, and fusion strategies.

\subsection{Modality-Specific QA Systems}
\begin{figure*}[h!]
  \centering
  \includegraphics[width=\textwidth]{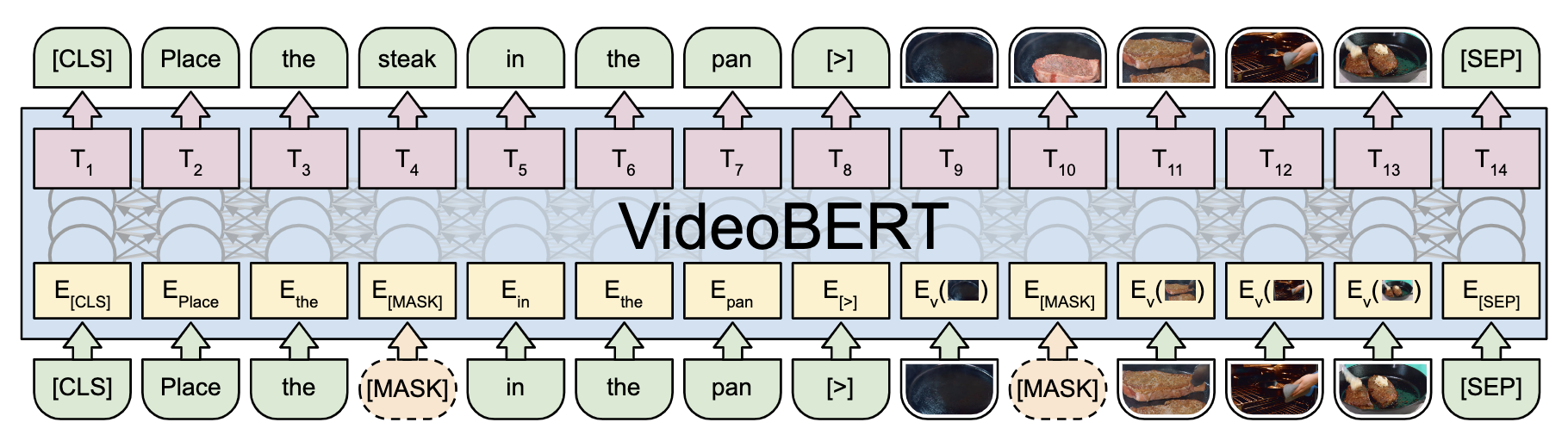}  
  \caption{VideoBERT jointly models text and video by learning cross-modal representations with masked token prediction across both modalities ~\cite{sun2019videobertjointmodelvideo}.}
  \label{fig:videoBert-architecture}
\end{figure*}

\begin{figure*}[ht!]
  \centering
  \includegraphics[width=0.95\textwidth]{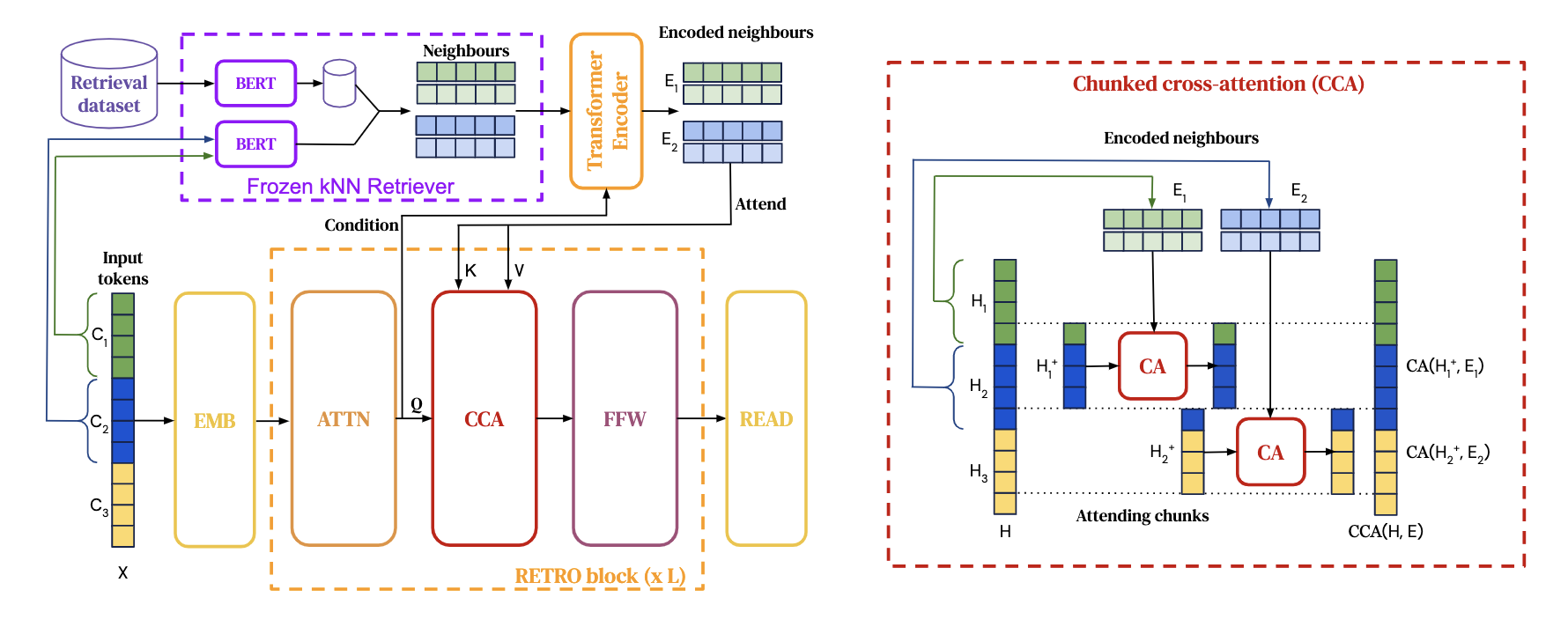}  
  \caption{\textbf{RETRO Architecture.} \textit{Left:} A simplified illustration where an input sequence of length \( n = 12 \) is divided into \( l = 3 \) chunks, each containing \( m = 4 \) tokens. For every chunk, \( k = 2 \) nearest-neighbor segments are retrieved, each consisting of \( r = 5 \) tokens. The retrieval pathway is depicted above the sequence. \textit{Right:} A closer view of the interactions within the CCA operator. Causal structure is preserved: the neighbors retrieved for the first chunk influence only the final token of that chunk and the tokens in the subsequent chunk~\cite{borgeaud2022improvinglanguagemodelsretrieving}.}
  \label{fig:retro-architecture}
\end{figure*}

\textbf{Unimodal Language QA (text only):} Recent advances in retrieval-augmented language models have significantly improved performance on open-domain question answering (QA) tasks. Traditional dense retrieval methods like DPR~\cite{karpukhin2020densepassageretrievalopendomain} and generative QA models such as Fusion-in-Decoder (FiD)~\cite{izacard2021leveragingpassageretrievalgenerative} have laid the groundwork for architectures that combine retrieval and generation. Building on this, \textbf{RETRO} (Retrieval Enhanced Transformer)~\cite{borgeaud2022improvinglanguagemodelsretrieving} ~\Cref{fig:retro-architecture} introduces a scalable approach that integrates nearest-neighbor retrieval into the transformer architecture using chunk level memory. Unlike traditional RAG style pipelines, RETRO retrieves from a large corpus during inference and feeds the retrieved chunks directly into the decoder, allowing a 7B model to match or exceed GPT-3 ~\cite{brown2020languagemodelsfewshotlearners} performance on knowledge-intensive tasks without requiring internet access or dynamic crawling.\\
Complementing this line of work, \textbf{ColBERTv2}~\cite{santhanam2022colbertv2effectiveefficientretrieval} addresses efficiency bottlenecks in dense retrieval through lightweight late interaction, enabling scalable retrieval over billions of passages. By decoupling encoding and interaction phases, ColBERTv2 delivers high throughput while preserving fine grained semantic matching making it highly suitable for open domain QA pipelines when paired with large language models. These methods outperform early dual-encoder systems by enabling fast and expressive ranking without sacrificing latency.\\
Further enhancements have come from methods like \textbf{Atlas}~\cite{izacard2022atlasfewshotlearningretrieval}, which fine-tune retrieval and generation jointly using few-shot learning; and \textbf{InPars}~\cite{bonifacio2022inparsdataaugmentationinformation}, which improves retriever performance using high quality synthetic QA pairs generated from instruction-tuned LLMs. Additionally, \textbf{BGE models}~\cite{bge_embedding} have become widely adopted for their strong zero-shot retrieval performance across \textbf{BEIR 2.0}~\cite{thakur2021beirheterogenousbenchmarkzeroshot}, demonstrating the importance of training general-purpose embedding models for knowledge-intensive QA.\\
\textbf{Static Vision-Language QA}:
Visual Question Answering (VQA) focuses on answering natural language questions based on visual input, typically an image. Early benchmark datasets such as VQA v2~\cite{jia2024vqa2visualquestionanswering}, VizWiz~\cite{gurari2018vizwizgrandchallengeanswering}, and GQA~\cite{ainslie2023gqatraininggeneralizedmultiquery} introduced grounded evaluation settings for visual linguistic reasoning. Traditional models relied on dual encoder or attention based fusion mechanisms over CNN and RNN representations~\cite{alomar2024rnnscnnstransformershuman}, but recent advances have shifted toward transformer-based architectures and vision language pretraining.\\
Modern VQA systems increasingly leverage multimodal transformers pretrained on large-scale corpora, such as LXMERT~\cite{tan2019lxmertlearningcrossmodalityencoder},\\UNITER~\cite{chen2020uniteruniversalimagetextrepresentation}, and VinVL~\cite{zhang2021vinvlrevisitingvisualrepresentations}, which enable fine-grained alignment of visual regions and language tokens. These models have been further surpassed by large scale foundation models like Flamingo~\cite{alayrac2022flamingo} and BLIP-2~\cite{li2023blip2bootstrappinglanguageimagepretraining}, which perform few-shot VQA via frozen vision encoders and language decoders, often outperforming finetuned baselines with minimal data. These advances suggest a paradigm shift from fully supervised fusion to general-purpose vision-language alignment with strong zero shot capabilities.\\
\textbf{Spatiotemporal Vision Language QA :} Video Question Answering (Video QA) involves answering natural language questions based on spatiotemporal visual input. Compared to image based QA, Video QA introduces additional challenges of temporal grounding, event localization, and multimodal synchronization (e.g., audio, motion, and subtitles). Foundational benchmarks such as \textbf{TVQA}~\cite{lei2019tvqalocalizedcompositionalvideo}, \textbf{HowTo100M}~\cite{miech2019howto100mlearningtextvideoembedding}, and \textbf{ActivityNet-QA}~\cite{yu2019activitynetqadatasetunderstandingcomplex} enabled early research on temporal alignment and multimodal feature fusion. Recent work explores transformer based architectures for modeling video sequences, such as \textbf{VideoBERT}~\cite{sun2019videobertjointmodelvideo}~\Cref{fig:videoBert-architecture} and \textbf{FrozenBiLM}~\cite{yang2022zeroshotvideoquestionanswering}, which leverage pretraining on large-scale instructional videos and pair vision features with text tokens. Multimodal pretrained models like \textbf{MERLOT Reserve}~\cite{zellers2021merlotmultimodalneuralscript} and \textbf{EgoVLP}~\cite{lin2022egocentricvideolanguagepretraining} achieve strong results by incorporating motion cues, subtitles, and egocentric views into unified encoders. \textbf{Ego4D QA}~\cite{di2024groundedquestionansweringlongegocentric} expands the domain to first person video understanding, evaluating temporal and action oriented reasoning through naturalistic tasks.
Formally, many video QA models treat the task as temporal answer grounding, aiming to select a time span \([t_s, t_e]\) within the video that is most relevant to the question \(q\):
\[
[t_s, t_e]^* = \arg\max_{[t_s, t_e]} \text{score}(q, V_{[t_s:t_e]})
\]
where \(V_{[t_s:t_e]}\) denotes the video segment and \(\text{score}(\cdot)\) is a learned multimodal matching function. Collectively, these developments reflect a transition from handcrafted feature fusion to large scale pretraining on instructional and egocentric videos, enabling better temporal reasoning and generalization across video based QA benchmarks.

\textbf{Acoustic-Language QA} focuses on answering questions from spoken content or environmental sounds, facing challenges such as temporal alignment, ASR errors, and noisy conditions. Benchmarks like \textbf{CLEAR}~\cite{lin2021clear} and \textbf{AVQA} ~\cite{phukan2024multilingualaudiovisualquestionanswering} extend beyond speech to include reasoning over non-speech audio and synchronized audio--visual streams. A key obstacle is ASR noise, especially in low-resource or noisy settings, addressed through robust self-supervised encoders (e.g., \textbf{wav2vec 2.0} ~\cite{baevski2020wav2vec20frameworkselfsupervised}, HuBERT), phonetic/subword retrieval, and cross-modal fusion. Modern models such as \textbf{SpeechT5} ~\cite{ao2022speecht5unifiedmodalencoderdecoderpretraining} and \textbf{Whisper} ~\cite{radford2021learning} enable multilingual QA and robust intent alignment. In low-resource contexts, domain-adaptive pretraining, pseudo-labeling, and contrastive noise-clean alignment improve performance, marking a shift from transcript-dependent approaches to direct audio--language understanding.

\subsection{Task-Oriented QA Systems}
\textbf{Modality-Aware Entity QA : }
Fact-based Question Answering focuses on retrieving concrete and objective information from a given context. These questions typically have a single correct answer, often grounded in explicit statements within the source material. In multimodal settings, fact-based QA involves extracting named entities, dates, attributes, or counts from text, images, or video transcripts. For example, in a video QA context, a fact based question might ask, ``What color is the car in the second scene?'' or ``How many people are standing near the counter?'' Models designed for this task prioritize precision and span-based extraction, often leveraging alignment between modalities and pretrained encoders for entity recognition and grounding~\cite{zhu2016visual7w, antol2015vqa, lei2018tvqa}.\\
\textbf{Causal Reasoning QA:} Explanatory Question Answering requires not only retrieving information but also performing complex reasoning, inference, and causal interpretation across one or more modalities. Unlike factoid QA, which often yields short span-based answers, explanatory QA demands structured, coherent responses that justify the answer through evidence synthesis and multihop reasoning. In multimodal scenarios, this involves integrating temporal video context, visual semantics, and spoken or written language to generate explanations. These systems often employ graph-based or transformer based reasoning modules to connect evidence across frames and modalities. Formally, explanatory QA can be framed as generating an answer \( a \) given a question \( q \) and context \( C = \{m_1, m_2, ..., m_k\} \) over multiple modalities, where the goal is to maximize:
\[
a^* = \arg\max_{a} \, P(a \mid q, C)
\]
where \( C \) includes multimodal inputs such as visual frames, audio transcriptions, and subtitle tokens. Datasets like VCR~\cite{zellers2019vcr}, HellaSwag~\cite{zellers2019hellaswag}, and HotpotQA~\cite{yang2018hotpotqa} have been pivotal in advancing this area by requiring models to reason about intent, causality, and implicit knowledge. Explanatory QA challenges models to move beyond pattern recognition, demanding fine-grained temporal alignment, causal chaining, and commonsense understanding in open-world settings.

\textbf{Contextual Interaction QA :} Conversational Question Answering involves maintaining multi-turn dialogue context to answer questions that depend on previous exchanges. Unlike standalone QA tasks, conversational QA systems must resolve coreference, ellipsis, and context-dependent queries. For example, given a conversation history, a user might ask, ``What did he say after the meeting?'' which requires linking ``he'' and ``the meeting'' to entities and events mentioned earlier. This task becomes even more complex in multimodal settings, where visual or audio cues from video must be aligned with the evolving dialogue. Effective conversational QA models integrate dialogue history, perform contextual grounding, and manage dialogue state to generate accurate and coherent responses~\cite{reddy2019coqa, choi2018quac, yang2022zeroshotvideoquestionanswering}.\\
\textbf{Temporal Event QA : }
Temporal or Event-based QA focuses on understanding the sequence, duration, and causality of events, particularly within dynamic modalities like video. This often involves identifying actions within specific time windows and modeling temporal dependencies. A key technique used is temporal attention over frame or segment-level features:

\[
\alpha_t = \frac{\exp\left(\mathbf{q}^\top \mathbf{k}_t\right)}{\sum_{t'} \exp\left(\mathbf{q}^\top \mathbf{k}_{t'}\right)}, \quad
\mathbf{v}_{\text{attn}} = \sum_t \alpha_t \cdot \mathbf{v}_t
\]

Here, \( \mathbf{q} \) is the question embedding, \( \mathbf{k}_t \) and \( \mathbf{v}_t \) are the key and value features at time step \( t \), and \( \mathbf{v}_{\text{attn}} \) is the temporally attended representation. This mechanism enables models to focus on relevant video segments to answer questions like “What happened after the person sat down?”

Models such as TVQA+~\cite{lei2020tvqa+} and HERO~\cite{li2020hero} leverage such techniques for robust temporal grounding in QA.\\
\textbf{Cross-modal Reasoning QA : }
Cross-modal Reasoning QA involves reasoning across multiple modalities e.g., vision, audio, and text requiring alignment and fusion of diverse input streams. A common approach is to use contrastive alignment losses to bring semantically related representations closer:

\[
\mathcal{L}_{\text{contrast}} = -\log \frac{\exp\left(\text{sim}(\mathbf{x}, \mathbf{y}^+)\right)}{\sum_{j} \exp\left(\text{sim}(\mathbf{x}, \mathbf{y}_j)\right)}
\]

where \( \text{sim}(\mathbf{x}, \mathbf{y}) \) is a similarity function (e.g., cosine similarity), \( \mathbf{x} \) is the question or text embedding, and \( \mathbf{y}^+ \) is the aligned video or image segment. This loss enforces cross-modal alignment critical for answering questions such as “What is the person doing while saying this?”

Furthermore, attention-based fusion is applied over different modality embeddings:

\[
\mathbf{z} = \text{MultiModalFusion}(\mathbf{x}_{\text{text}}, \mathbf{x}_{\text{video}}, \mathbf{x}_{\text{audio}})
\]
This fused representation \( \mathbf{z} \) is then used for downstream QA prediction. Models like Flamingo~\cite{alayrac2022flamingo} and JustAsk~\cite{yang2021justask} adopt such mechanisms to achieve State of the art performance on complex, multimodal QA tasks.

\section{Multimodal Retrieval Strategies for QA Systems}
\label{sec:retrieval_mechanisms}
Multimedia QA systems rely on accurate, modality aware retrieval mechanisms to locate relevant data segments across textual, visual, audio, and video modalities. Below, we explore five key retrieval paradigms that underpin these systems.




\subsection{Dense Retrieval}
Dense retrieval systems have emerged as a powerful alternative to traditional lexical matching techniques like BM25, particularly in open-domain question answering and information retrieval tasks ~\cite{karpukhin2020densepassageretrievalopendomain}. These approaches embed both queries and documents into a shared semantic space using deep neural encoders, allowing them to capture latent semantic relationships and perform soft matching beyond exact token overlaps. A key advantage of dense retrieval is its ability to handle vocabulary mismatch and contextual nuances, which are often problematic for sparse vector models.

Let \( f(q) \) and \( g(d) \) denote the vector representations of a query \( q \) and a document \( d \), respectively, as produced by their respective encoders. The similarity score between a query and document is typically computed using an inner product:

\[
\text{score}(q, d) = f(q)^\top g(d)
\]

Training such models often relies on contrastive learning, where the model learns to distinguish between relevant and irrelevant document-query pairs. Given a positive document \( d^+ \) and a set of negatives \( \{d^-\} \), the contrastive loss can be expressed as:

\[
\mathcal{L}_{\text{contrastive}} = -\log \frac{\exp(\text{score}(q, d^+))}{\exp(\text{score}(q, d^+)) + \sum_{d^-} \exp(\text{score}(q, d^-))}
\]

One of the pioneering systems in this space is Dense Passage Retrieval (DPR)~\cite{karpukhin2020dense}, which uses dual BERT encoders, one for questions and one for passages, and trains them on a large corpus of question answer pairs. DPR showed strong performance on benchmarks like Natural Questions and TriviaQA, outperforming sparse methods in recall oriented settings. However, dual encoder models are sometimes limited by their coarse-grained similarity function.

To address this, ColBERT~\cite{khattab2020colbert} introduced a late interaction mechanism that computes token-level similarity between query and document embeddings. Each query token \( q_i \) is matched to its most similar document token \( d_j \), and the final score aggregates the maximum similarities:

\[
\text{score}_{\text{ColBERT}}(q, d) = \sum_{i} \max_j \cos(q_i, d_j)
\]

This formulation allows ColBERT to retain finegrained semantic matching while remaining efficient via pre indexed document representations. More recent works, such as GTR~\cite{ni2021largedualencodersgeneralizable} and RocketQA~\cite{qu2021rocketqaoptimizedtrainingapproach}, further improve dense retrieval by incorporating multi view learning, hard negative mining, and advanced distillation techniques. Despite their success, dense retrieval models often face challenges in training stability, negative sampling strategies, and zero-shot generalization, making this an active area of research.

\subsection{Embedding Retrieval}
Multimodal retrieval embeds diverse data types such as text, images, audio, and video into a shared latent space, enabling cross modal retrieval where a query in one modality can retrieve semantically aligned content in another. The primary challenge lies in learning unified representations across modalities that differ in structure, dimensionality, and temporal characteristics. Models like CLIP~\cite{radford2021learning} adopt a dual encoder architecture, where visual and textual inputs are processed independently and trained with a symmetric contrastive objective based on the InfoNCE loss:
\[
\mathcal{L}_{\text{contrast}} = -\log \frac{\exp(\text{sim}(\mathbf{v}_i, \mathbf{t}_i)/\tau)}{\sum_j \exp(\text{sim}(\mathbf{v}_i, \mathbf{t}_j)/\tau)}
\]
Here, $\text{sim}(\cdot)$ denotes cosine similarity and $\tau$ is a temperature parameter controlling the sharpness of the distribution. While CLIP focuses purely on contrastive alignment, models like BLIP~\cite{li2022blip} enhance flexibility by integrating both contrastive and generative objectives using a unified encoder decoder framework. This allows simultaneous optimization for retrieval and caption generation. Retrieval-augmented models such as Flamingo~\cite{alayrac2022flamingo} further incorporate few-shot capabilities by combining pretrained vision and language backbones with cross-attention layers, enabling dynamic fusion of multimodal context during inference. Recent work like ImageBind~\cite{girdhar2023imagebindembeddingspacebind} extends these ideas to six or more modalities, including depth, thermal, and audio, using a single encoder to embed all modalities into a common space. Additionally, techniques such as hard negative mining, modality dropout, and curriculum learning are being actively explored to enhance alignment quality, improve sample efficiency, and boost performance in zero shot and openset scenarios.

\subsection{Cross-Modal Retrieval}
Cross-modal retrieval refers to using inputs from one modality to retrieve another such as querying with text to retrieve videos. This setting requires asymmetric mappings between modalities and often employs late fusion strategies or co-attention mechanisms.

A typical scoring function for cross-modal retrieval can be represented as:
\[
\text{score}(q, v) = \text{cos}(\phi_T(q), \phi_V(v))
\]
where \(\phi_T\) and \(\phi_V\) are projection functions for text and visual inputs.\\
Advanced systems such as VATT~\cite{akbari2021vatttransformersmultimodalselfsupervised} and MMT~\cite{gabeur2020multi} use self supervised training to ensure alignment and discriminative representations. These models leverage transformer based backbones with fusion layers to capture inter modality relationships and finegrained temporal cues.

\subsection{Temporal Video Segment Retrieval}
\begin{figure}[h!]
  \centering
  \includegraphics[width=0.90\columnwidth]{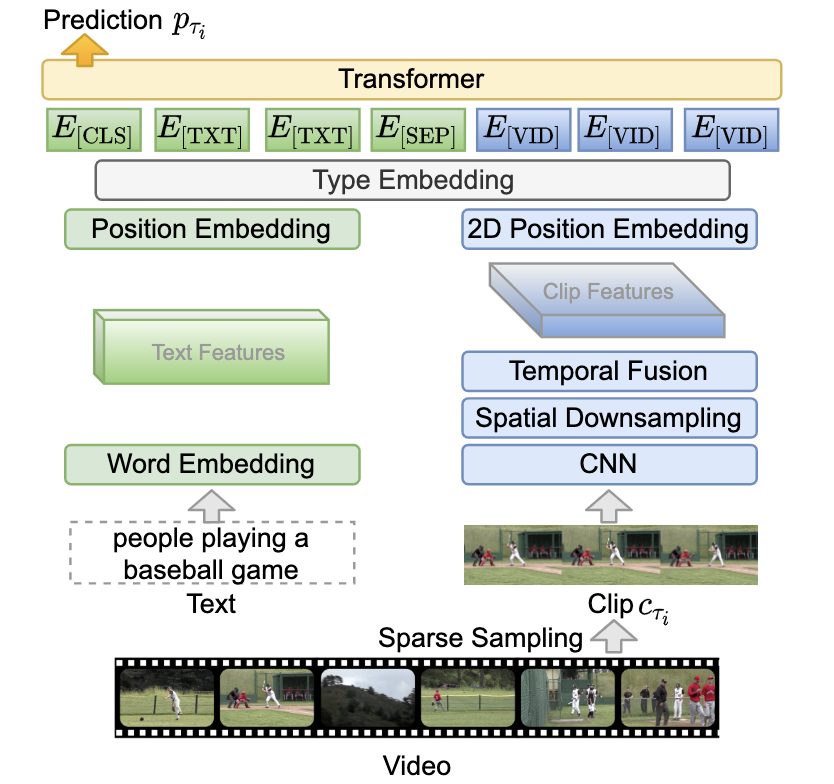}  
  \caption{Overview of the CLIPBERT architecture. The diagram illustrates prediction for a single sampled clip. When multiple clips are sampled, their individual predictions are aggregated to produce the final result.~\cite{lei2021less}.}
  \label{fig:clipBert}
\end{figure}

Temporal retrieval aims to identify the most semantically relevant segment of a video in response to a natural language query. This task is commonly framed as temporal grounding or span prediction, where a video \(V\) and query \(q\) are given, and the goal is to retrieve the optimal time interval \([t_s, t_e]\) that maximizes alignment with the query:
\[
[t_s, t_e]^* = \arg\max_{[t_s, t_e]} \text{score}(q, V_{[t_s:t_e]})
\]
Here, \(\text{score}(\cdot)\) denotes a learned relevance function, often parameterized by a multimodal encoder. Models like HERO~\cite{li2020hero} utilize hierarchical transformers to encode both global video context and fine-grained clip level information, integrating temporal attention mechanisms to align sequential visual embeddings with language representations. Other methods, such as ClipBERT~\cite{lei2021less} ~\Cref{fig:clipBert}, optimize for computational efficiency by employing sparse temporal sampling and late fusion of visual language features, allowing scalable training without processing entire video sequences. These systems often incorporate pretrained vision language models, self attention over frame query pairs, and auxiliary losses like frame level alignment or contrastive span ranking to improve localization accuracy. Recent trends also explore multimodal fusion via temporal cross attention and query-aware temporal pooling to better capture long range dependencies and subtle temporal cues across diverse video content.

\subsection{Audio-Visual Retrieval}
Audio-visual retrieval involves learning joint representations from temporally aligned audio and visual signals, enabling cross modal search tasks such as speaker localization, event detection, and scene understanding. State of the art approaches like AVTS~\cite{korbar2018cooperative} and AVID~\cite{morgado2020audio} leverage large scale unlabeled videos to learn self-supervised embeddings by maximizing the correspondence between audio and visual inputs. These models typically incorporate 2D or 3D convolutional networks for video encoding and log mel spectrogram encoders for audio, followed by fusion modules that use cross modal attention or late fusion strategies to integrate both streams. Synchronization plays a crucial role; hence, techniques often employ temporal contrastive objectives that ensure temporally coherent frames and audio segments are mapped to nearby points in the embedding space. To maintain temporal granularity, temporal convolutions and dilated attention mechanisms are used, while projection heads align embeddings into a shared latent space suitable for retrieval. These representations enable flexible retrieval scenarios retrieving audio based on visual cues, or vice versa and serve as robust backbones in downstream tasks such as audio visual question answering and multimodal summarization.
\section{Multimodal QA Architectures and Benchmarks}

Modern Multimodal QA systems are underpinned by architectural frameworks that must efficiently align, represent, and reason over heterogeneous modalities text, vision, audio, and video each with distinct temporal, spatial, and semantic characteristics. Four dominant design paradigms have emerged, each addressing different modeling and system-level trade-offs.\\
The \textbf{Retrieve then Read} paradigm decouples retrieval and reasoning through a two-stage pipeline. Dense retrievers, often based on dual-encoder architectures like CLIP~\cite{radford2021learning}, BLIP-2~\cite{li2023blip}, or Video-MAE~\cite{tong2022videomae}, compute similarity between query and content embeddings, while sparse retrievers may use keyword matching over transcriptions or OCR. Retrieved multimedia elements (e.g., keyframes, subtitles, motion features) are encoded using frozen or fine-tuned modality-specific backbones. A typical dense retrieval objective can be formulated using a contrastive loss:
\[
\mathcal{L}_{\text{contrastive}} = -\log \frac{\exp(\text{sim}(q, d^+))}{\sum\limits_{d \in \mathcal{D}} \exp(\text{sim}(q, d))}
\]
where \( q \) is the query embedding, \( d^+ \) is the positive document, and \( \mathcal{D} \) includes both positive and negative candidates. This architecture promotes modularity, facilitates offline indexing and caching, and scales well for large corpora. However, it often lacks tight alignment between modalities and struggles with resolving temporal dependencies or cross-modal co reference at fine granularity~\cite{zhu2023minigpt4}.\\
In contrast, \textbf{End-to-End Fusion} models directly encode multimodal inputs using shared or hybrid encoders. Early fusion concatenates raw or low level embeddings across modalities and feeds them to a single encoder, whereas mid-level fusion introduces modality-specific encoders with interaction layers such as multi head cross-attention to enable alignment. Late fusion strategies maintain modality specific pipelines until final integration, often using gated summation or attention based pooling. These models are frequently implemented using Transformer variants like ViLT~\cite{kim2021vilt}, FLAVA~\cite{singh2022flava}, Unified-IO~\cite{lu2023unifiedio}, or more recently MM-ReAct~\cite{yang2023mmreact}. End-to-end fusion enhances joint reasoning and fine-grained alignment, but at the cost of scalability and compute efficiency.\\
The \textbf{LLM + Multimodal Retriever} class extends retrieval-augmented generation to multimodal contexts. Instruction tuned language models (e.g., GPT-4V~\cite{openai2023gpt4v}, LLaVA~\cite{liu2023llava}, or Gemini~\cite{google2023gemini}) are paired with modality-aware retrievers that operate over pre-indexed visual, auditory, or video content. Examples include Video-RAG~\cite{seo2024videorag}, RETRO~\cite{borgeaud2022improving}, and MM-ReAct~\cite{yang2023mmreact}, where queries are formatted as prompts that guide retrieval and condition the LLM’s generation. These architectures enable explainability, compositional reasoning, and integration of retrieved external knowledge, while maintaining flexibility through in-context learning. However, they rely heavily on retrieval quality and alignment between retrieved content and prompt structure.\\
Finally, \textbf{Knowledge-Grounded Multimodal QA} architectures incorporate structured external information such as scene graphs, audio event graphs, spatial temporal interaction graphs, or commonsense knowledge bases like ConceptNet~\cite{speer2017conceptnet} or ATOMIC~\cite{sap2019atomic}—to guide reasoning. These systems often use graph neural networks (GNNs), memory augmented transformers, or retrieval-enhanced modules to encode and query structured knowledge aligned with visual or auditory streams. This grounding improves factual correctness, enables multi-hop inference, and supports counterfactual or causal reasoning~\cite{xu2023vidskg}, though it introduces additional complexity in knowledge extraction and alignment.


\section{Conclusion}
Multimodal Question Answering is undergoing a transformative shift through the integration of large scale multimedia retrieval systems. By leveraging text, image, video, and audio sources, modern QA pipelines are moving beyond static knowledge toward contextually rich, temporally grounded, and semantically aligned responses. Despite recent progress, several challenges remain unresolved. Key issues include the difficulty of finegrained multimodal alignment (e.g., syncing spoken language with visual scenes), the lack of robust trustworthiness mechanisms such as modality attribution or segment-level citations, and the computational overhead introduced by real time or large scale retrieval. Further complexities arise in handling multilingual queries and supporting low-resource modalities, along with the persistent challenge of evaluating answer quality across modalities.\\
Addressing these limitations opens several promising research directions. One is the development of \emph{multimodal retrieval augmented generation (RAG)} systems that provide transparent explanations and evidence. Another is the push toward \emph{unified embedding spaces} for efficient and scalable cross modal retrieval. Future systems must also prioritize \emph{lightweight architectures} for deployment in resource-constrained environments, \emph{promptable retrievers} that adapt dynamically to evolving multimedia content, and real time QA pipelines capable of understanding \emph{live-streamed data} such as meetings, surveillance footage, and egocentric videos.\\
To catalyze progress, the community must invest in standardized benchmarks, open source toolkits, and shared evaluation protocols. Equally important is the commitment to building QA systems that are not only accurate but also interpretable, trustworthy, and responsive across real world multimedia settings.


\bibliographystyle{ACM-Reference-Format}
\balance
\bibliography{main}
\nocite{aiqam25}

\end{document}